\documentclass[prl,preprint,showpacs,preprintnumbers,amsmath,amssymb]{revtex4}
\usepackage{graphicx}
\usepackage{dcolumn}
\usepackage{bm}

\begin{document}
\draft
\title{ Field-induced Ferromagnetic Order and Colossal Magneto Resistance in  La$_{1.2}$Sr$_{1.8}$Mn$_2$O$_7$: a $^{139}$La NMR study}
\author{Y. Shiotani$^{1, \dagger}$,  J. L. Sarrao$^2$, and Guo-qing Zheng $^{1,3,}$}\altaffiliation[author to whom correspondence should be addressed. ]{e-mail: zheng@psun.phys.okayama-u.ac.jp}
\address{$^1$Department of Physical Science, Graduate School of Engineering
Science, Osaka University, Osaka 560-8531, Japan}
\address{$^2$ Condensed Matter and Thermal Physics, MS K764, Los Alamos National Laboratory, Los Alamos, NM 87545, USA}
\address{$^3$ Department of Physics, Okayama University, Okayama 700-8530, Japan}
\date{\today}

\begin{abstract}
In order to gain insights into the origin of colossal magneto-resistance (CMR) in manganese oxides, we performed a $^{139}$La NMR study in the double-layered compound La$_{1.2}$Sr$_{1.8}$Mn$_2$O$_7$. We find that above the Curie temperature $T_C$=126 K, applying a magnetic field induces a {\it long-range} ferromagnetic order that persists up to $T$=330 K. The critical field at which the induced magnetic moment is saturated coincides  with the field at which the CMR effect reaches to a maximum. Our results therefore indicate that the CMR observed above $T_C$ in this compound is due to the field-induced ferromagnetism that produces a metallic state via the double exchange interaction.

\end{abstract}

\maketitle
\widetext


Since the discovery of giant magneto-resistance (GMR) in magnetic superlattices,  spin-dependent conductivity  has become an important  subfield of research in both condensed matter and applied physics. In recent years, the Ruddlesden-Popper series of manganese oxides, (La,Sr)$_{n+1}$Mn$_n$O$_{3n+1}$,   has been found to show even larger magnetoresistance, the so-called colossal  magneto-resistance (CMR) \cite{Jin,Moritomo,Kimura,Dagotto0}. These compounds have  perovskite crystal structure that also houses high temperature superconductivity. It has become clear that in these compounds, the degrees of  spin, charge, electronic orbital and lattice interact with each other, and are responsible for the rich physical properties \cite{Tokura}. 

Among them,  the $n$=2 compounds, La$_{2-2x}$Sr$_{1+2x}$Mn$_2$O$_7$ have  attracted a great deal of attention. The carrier content (hole content) can be systematically tuned  by changing $x$. Correspondingly, the magnetic structure and the electrical conductivity change \cite{Moritomo,Kimura}. The $c$-axis  length is about 5-fold larger than the $a$-axis; the ferromagnetic   MnO$_2$ double layers and non-magnetic, insulating (La,Sr)$_2$O$_2$ block are stacked along the $c$-axis, forming a quasi two dimensional structure. 
This composes a natural ferromagnetic-nonmagnetic-ferromagnetic (FM/NM/FM) superlattice.  The resistivity along the $c$-axis, $\rho_c$, through a tunneling magnetoresistance (TMR) process, is larger than the resistivity along the $a$-axis, $\rho_a$, by a factor of 10$^3\sim$ 10$^4$ \cite{Moritomo}.  
The magnetic properties are complicated, depending on the coupling between adjacent MnO$_2$ double layers. For $x\leq$0.48, the ground state is a ferromagnetically ordered state. For larger $x$, however, the low temperature state is an antiferromagnetically ordered state \cite{Perring,Kubota,Kubota2}.

The most extensively investigated composition is the $x$=0.4 compound which orders ferromagnetically at $T_C$=126 K and shows a negative magneto-resistance of 2$\times$ 10$^4$ \% around $T_C$ \cite{Moritomo,Kimura}. Intriguingly, this CMR effect persists in the paramagnetic state up to $T\sim$300 K \cite{Moritomo}.   The most straightforward, existing interpretation for the CMR  might be through the double exchange interaction theory \cite{Furukawa}. Namely, applying a magnetic field aligns the Mn spins of the three electrons in the $t_{2g}$ orbit, which in turn forces the spin in the $e_{g}$ orbit to align in the same direction through Hund coupling so that the $e_{g}$ electron can hop to the neighboring site with minimal energy cost. However, this mechanism cannot explain the CMR effect above $T_C$ where a small field  brings about a huge magneoresistance effect \cite{Moritomo}. Thus, 
 the fundamental question of what is the origin of the CMR in this series of compounds remains unanswered. So far, several classes of theories have been proposed, which include  the "Random resistor network" theory \cite{Dagotto} on the basis of phase separation suggested by neutron scattering \cite{Perring}, polaron effect \cite{Millis} and the multi-critical point scenario \cite{Nagaosa}.

In this Letter, we report a $^{139}$La NMR (nuclear magnetic resonance) study in La$_{2-2x}$Sr$_{1+2x}$Mn$_2$O$_7$ ($x$=0.4),  focusing on the relationship between the magnetic properties and the occurrence of CMR. There are two crystallographically different sites of La; one lies in the middle of the MnO$_2$ bilayer ( hereafter we will call this site in-bilayer site), and the other outside the MnO$_2$ bilayer (out-bilayer site). Therefore, La NMR is a powerful tool to probe the magnetic state of the  MnO$_2$ bilayer. 
Firstly, if Mn is in the ferromagnetic state, an NMR signal enhancement effect that is unique to such state will be observed. Secondly, 
 the in-bilayer La site will experience a larger internal field than the out-bilayer site. In contrast, if the  MnO$_2$ bilayer is in an antiferromagnetic state, then the internal field at the in-bilayer La site  will cancel.

We found, unprecedentedly, that applying a magnetic field induces  ferromagnetism above $T_C$, which is primarily responsible for the CMR. The field induces a long-range ferromagnetic order, which drives the system metallic via the double exchange interaction. 



Single crystal  of La$_{1.2}$Sr$_{1.8}$Mn$_2$O$_7$ used in this study
was grown by the traveling solvent floating zone (TSFZ)  method.  
X-ray diffraction indicates that the $c$-axis length is 20.20$\AA$.
For NMR measurements, the single crystal  was crushed into  a powder with
 particle size of $\sim$20$\mu$m to allow a maximal penetration of the
oscillating (RF) magnetic field, $H_1$, since the NMR signal intensity is proportional to $H_1^2$.
NMR experiments were
performed using a home-built phase-coherent spectrometer. A
standard $\pi$/2-$\tau$-$\pi$-echo pulse sequence was used.  The NMR spectra at zero magnetic field were taken by changing the RF frequency and recording the echo intensity step by step. The spectra at finite magnetic fields were taken by sweeping the field and recording the echo intensity with the aid of a Box-car integrator.


\begin{figure}
\begin{center}
\includegraphics[scale=0.6]{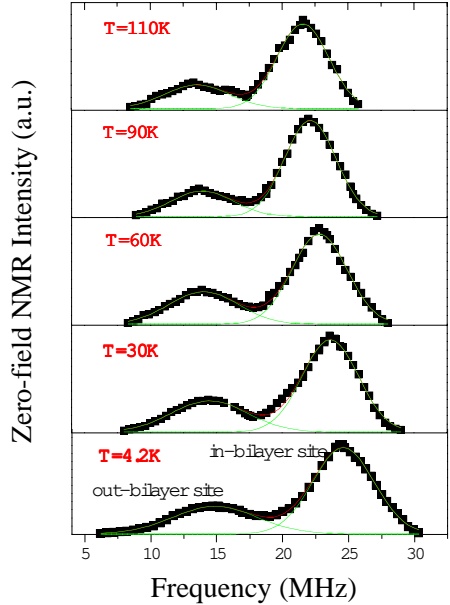}
\caption{(Color online) $^{139}$La NMR line shape at zero magnetic field in La$_{1.2}$Sr$_{1.8}$Mn$_2$O$_7$. The red curve is the sum  of two Gaussian functions (green) that best fit  the data.}
\label{fig:1}
\end{center}
\end{figure}

Figure 1 shows the NMR spectra at zero external magnetic field for various temperatures. Two peaks are observed. The signals were observed in a condition of strong $H_1$ enhancement. The pulse condition is $\pi$/2-$\tau$-$\pi$=1.0$\mu$s-10$\mu$s-2$\mu$s with RF pulses of a power of 5 mW which is a 10$^3$$\sim$10$^4$-fold smaller power than requested for paramagnetic materials.  This $H_1$ enhancement indicates that the signals come from the domain walls  in a ferromagnetic state. In a domain wall of a ferromagnet, the RF field acting on the nucleus, which is perpendicular to the static magnetic field, is much larger than the applied $H_1$ because of  the oscillating electron magnetic moment \cite{H1enhance}.
We have confirmed that the $H_1$ enhancement effect disappears and the signal to noise ratio decreases at high magnetic fields (above 0.4 T at $T$=4.2 K) since there the magnetic domain walls are removed.
Then, the peak seen at higher frequency is due to the La in-bilayer site. 

\begin{figure}
\begin{center}
\includegraphics[scale=0.6]{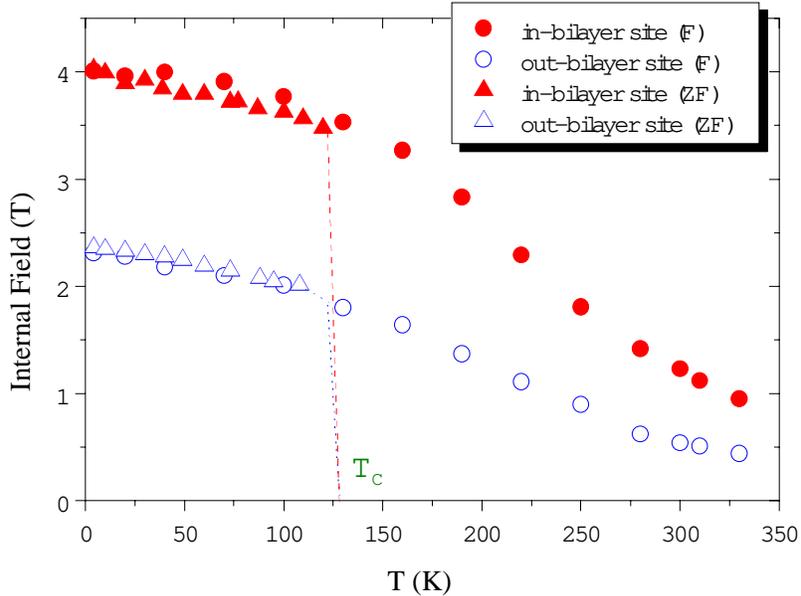}
\caption{(Color online) Temperature dependence of the internal magnetic field at the La sites in La$_{1.2}$Sr$_{1.8}$Mn$_2$O$_7$. The triangles are the data obtained by zero field (ZF) NMR. The circles are the data obtained by applying a finite magnetic field (F) of about 10 T (see text). The ZF Curie temperature is $T_C$=126 K.}
\label{fig:2}
\end{center}
\end{figure}

Figure 2 shows the temperature dependence of the internal field $H_{in}$ extracted from the spectra, according to $\omega=\gamma H_{in}$, where $\omega$ is the Larmor (RF) frequency, and $\gamma$ is the gyromagnetic ratio which is 6.0146 MHz/T for $^{139}$La. The large internal field is  due mainly to the transferred hyperfine field from the ordered spins at Mn 3d orbitals. The signal   has enough intensity at $T$=120 K, but is lost  above this temperature. This indicates that the Curie temperature $T_c$ is around this temperature, in good agreement with the value $T_C$=126 K reported  from neutron scattering measurements \cite{Perring,Kubota,Kubota2}. Also, our result suggests that the ferromagnetic transition is of first order phase transition, although no appreciable hysteresis with respect to temperature was found around $T_C$. The first-order like phase transition is similar to the cases of 
La$_{1-x}$Ca$_{x}$MnO$_3$ \cite{LaMnO3,Adams}, and La$_{1-x}$Na$_{x}$MnO$_3$ \cite{LaNaMnO3}.

The magnitude of the internal field is also confirmed by the NMR measurements under finite external fields.  Figure 3 shows the NMR spectra at Larmor frequency of 91.2 MHz. By utilizing the strong magnetization ($M$) anisotropy that $M_{ab}$ is larger by one order of magnitude than $M_{c}$ \cite{Moritomo}, mechanical shocks were provided when the sample is in the magnetic field of 12$\sim$16 T so that the $ab$-plane of the small grains align preferably to the  direction of the external field. Therefore the spectra can be regarded as being under the condition of $H\parallel ab$-plane.
The broken line in Fig. 3 indicates the resonance field in the absence of an internal field. The distance between the actually observed peak and the broken line is  therefore the internal field. As can been seen in Fig. 2, such extracted internal field below $T_C$ agrees excellently well with that obtained from zero field NMR measurements. Note that, at low temperatures, there is no appreciable signal peak at the position of the broaken line in Fig. 3. This is in contrast to the case   observed in the $x$=0.5 compound \cite{Shiotani}, where there appears a peak around this position, due to the in-bilayer La site that is in the antiferromagnetically ordered state. 

\begin{figure}
\begin{center}
\includegraphics[scale=0.6]{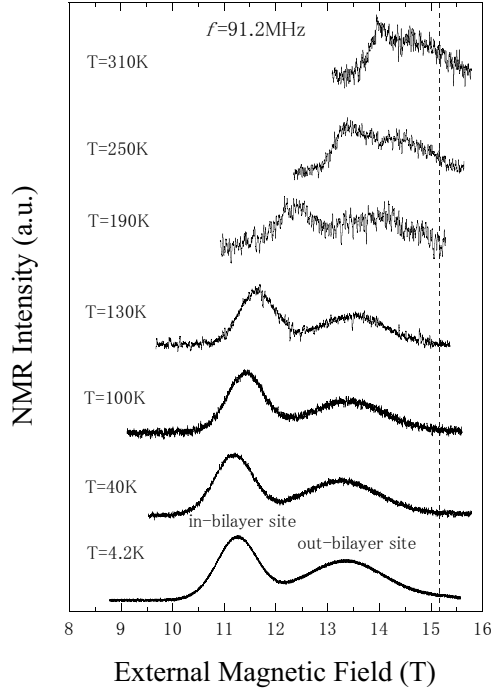}
\caption{Typical examples of the magnetic-field swept $^{139}$La NMR line shape at the Larmor frequency of $\omega_L$=91.2 MHz at various temperatures. The vertical broken line is the resonance field position $\omega_L/\gamma$, in the absence of internal field.}
\label{fig:3}
\end{center}
\end{figure}

\begin{figure}
\begin{center}
\includegraphics[scale=.6]{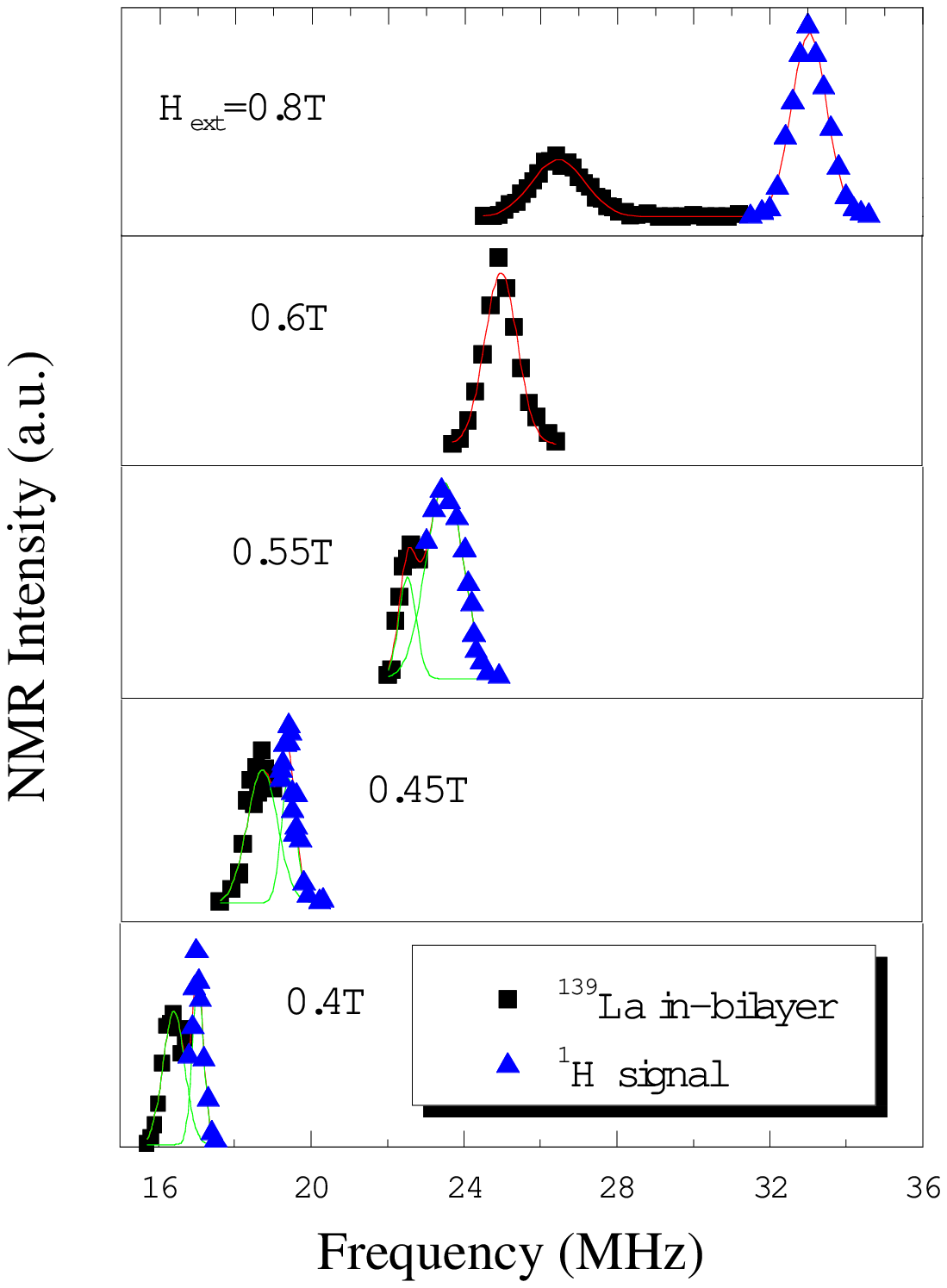}
\caption{(Color online) Examples of $^{139}$La  NMR spectra for the in-bilayer La site at $T$=130 K, in the presence of external magnetic fields, $H_{ext}$. The triangles indicate proton ($^1$H) spectrum, which, at $H_{ext}$=0.6 T, coincides with $^{139}$La spectrum. The red curve is the sum  of two Gaussian functions (green) that best fit  the data.}
\label{fig:4}
\end{center}
\end{figure}

The most important finding of this work  is that even above $T_C$, the internal field persists when an external field is present, as can be seen in Fig. 3. This means that ferromagnetism is induced by the applied field, which persists up to 330 K. This is the key to understanding the CMR effect above $T_C$ in this compound.

The saturated magnetic moment at the critical external field, $H_{cr}$, at which the moment is fully polarized depends on temperature.  Figure 4 shows an example of the spectra at $T$=130 K at various external fields. 
Below $H_{cr}$=0.58 T, the peak frequency $\omega_p$ increases with increasing field $H_{ext}$ with $\partial \omega_p/\partial H_{ext}$ $>$ $\gamma$, indicating significant contribution of the polarized Mn moment. Above $H_{cr}$, however, $\omega_p=\omega_0+\gamma H_{ext}$, where  $\omega_0$ is a constant. This indicates that above $H_{cr}$, the induced moment reaches to a fully polarized value.
Figure 5 shows the field induced moment as a function of external magnetic field at different temperatures. Here, we have assumed that at $T$=4.2 K, the saturated moment is 3.6$\mu_B$ \cite{Moritomo}.
At $T$=4.2 K, we have confirmed that the external field has no effect on the ordered  moment  because the moment is already fully polarized in the Curie state.   Note that $H_{cr}$ at 130 K and 150 K is 0.58 T and 2.93 T, respectively. These values agree excellently well with the $H_{cr}^{\rho}$ at which the magneto-resistance effect reaches to 90\% of  the maximal value \cite{Moritomo}. At still higher temperatures, $H_{cr}$ increases further, roughly tracking $H_{cr}^{\rho}$ \cite{note}.
Therefore, our results indicate that the CMR in this compound is due to the field-induced magnetism that makes the compound metallic via the double exchange interaction.

Finally, it is an interesting future work to clarify the origin of the field induced magnetism above $T_C$. A possible candidate might be that the susceptibility is highly enhanced due to competing interactions (double exchange interactions that favor ferromagnetic state and superexchange state that favors an antiferromagnetic order), or due to a proximity to a  critical point in the phase diagram of $T$ vs $x$ \cite{Nagaosa}. The tendency  of short-range order above $T_C$ in the absence of external field suggested previously \cite{Simon,Berger} is probably consistent with the former case.
For the later case, Murakami and Nagaosa have recently shown that the fluctuations are largely enhanced around a multiple critical point \cite{Nagaosa}, which gives an account for the first order ferromagnetic phase transition and the CMR effect in Re$_{1-x}$Sr$_x$MnO$_3$ (Re=La, Pr, Nd). 
We hope that our work will stimulate more theoretical studies in this regard.

\begin{figure}
\begin{center}
\includegraphics[scale=0.6]{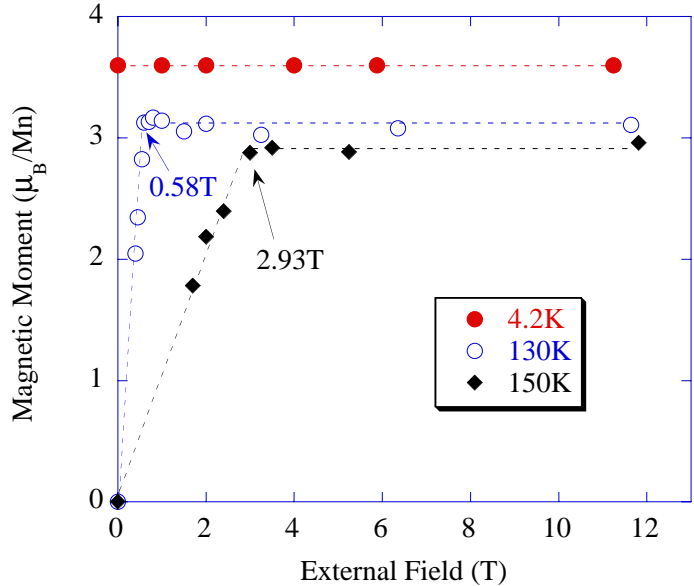}
\caption{(Color online) Magnetic moment of Mn estimated from the NMR frequency (see text) as a function of field at different temperatures. The broken lines are guides to the eyes. The arrow indicates $H_{cr}$ at $T$=130 and 150 K, respectively.}
\label{fig:5}
\end{center}
\end{figure}



In conclusion, we found that an external magnetic field induces a long-range ferromagnetic order above zero-field Curie temperature $T_C$=126 K. The critical field at which the induced magnetic moment is saturated agrees excellently well with the field at which the CMR effect reaches to a maximum. Our results indicate that the CMR observed above $T_C$ in this compound is due to the field-induced ferromagnetism that produces a metallic state via double exchange interaction.


We thank H. Saeki for X-ray diffraction measurement and Y. Kitaoka for  helpful discussion and comments. This work was  partially supported by  MEXT grants for scientific research.  Work at Los Alamos was performed under the auspices of the  DOE.

\vspace{1cm}

$\dagger$ Present address: RICOH Co. Ltd., Shin-Yokohama 3-2-3, Minato-Kita-ku, Yokohama 222-0033, Japan.


\begin{references}
\bibitem{note}
At higher temperatures, it becomes difficult to precisely define $H_{cr}$.
\end{references}
\end{document}